\documentclass{article} 
\usepackage{amssymb,amsmath,epsfig,cite}
\usepackage{sublabel}
\nonstopmode

\title{A mathematical model for a gaming community}

\author{Romulus Breban}
\date{\begin{center}
{\it Institut Pasteur, 75724 Paris Cedex 15, France}
\end{center}\today}

\begin{document} 
\maketitle 

\begin{abstract}
We consider a large community of individuals who mix strongly and meet in pairs to bet on a coin toss. We investigate the asset distribution of the players involved in this zero-sum repeated game. Our main result is that the asset distribution converges to the exponential distribution, irrespective of the size of the bet, as long as players can never go bankrupt. Analytical results suggests that the exponential distribution is a stable fixed point for this zero-sum repreated game. This is confirmed in numerical experiments.
\end{abstract}
\section{Introduction}

Communities of ranked players occur in a variety of situations. Casinos routinely accommodate live player communities engaged in gambling activities. Casino players may be ranked by the total value of their casino tokens. Traditional board (e.g., chess, backgammon, go, etc.) and card (e.g., bridge) games also bring together gaming communities centered around gaming federations. In this case, players are ranked by strength of play; e.g., the Elo system. With the advent of the Internet, society experienced, like never before, an increase in the number of gaming communities. First, the Internet has strengthen existent gaming communities. Second, Massive Multiplayer On-line Games (MMOG) of various type (e.g., role-playing, real-time or turn-based strategy, simulations of sports or racing events, etc.) were made available to host gaming communities in cyberspace. Popular MMOGs have been: Ultima Online, EverQuest and World of Warcraft, to name a few. Nearly all MMOGs use virtual currency that players earn and spend during the game. Hence, players are easily ranked by the amount of virtual currency.

Generally, the gaming experience of one player is divided into a large numbers of matches, which are single player, multiplayer or, possibly, team encounters, where the number of players involved is much smaller than the community size. As a result of playing, individuals acquire gaming experience and their ranks change actively. This setup, known as an iterated or repeated game \cite{Mertens:1994ve}, has been extensively applied to the prisoner's dilemma \cite{Hamilton:1981vw,Kaznatcheev:2015}, as well as other dilemmas \cite{Capraro:2013dz}. The focus of previous work has been primarily finding game strategies which have good return \cite{Hamilton:1981vw,Nowak:1993rm,Press:2012jh,Hilbe:2013ip,Wang:2016db,Capraro:2013dz} and differentiating traits of these strategies \cite{Nowak:1993rm,Hilbe:2013ip,Wang:2016db}. Repeated zero-sum games have received substantially less atention \cite{Rosenberg:2001ju,Lehrer:2010iu,Sorin:2011bj,Akian:2015,ZiliottoB:2016,Carmona:2016s}.

In this paper, we develop a model for a community of players who mix strongly and meet in random pairs to play one-on-one matches. We assume that, as a result, their ranks change according to the rule of a zero-sum game. Hence, game rank does not stand for strength of play and may be more approriately regarded as a game asset.  This setup is equivalent to that of a community of players who play their assets according to a zero-sum game. The zero-sum postulate guarantees that the assets (or ranks) of all players combined are conserved, and provide the possibility for a stationary distribution of assets to exist. Assuming that players play such that they avoid bankruptcy, we study the dynamics of asset distribution, using both analytical and computational tools. 

\section{Modeling framework}
We cosider $N$ individuals and denote their assets by $a_i$, $i=1,..., N$. We assume that individuals pair up randomly, irrespective of game rank, to play matches of a zero-sum game. Say, player $i$ meeets player $j$, where $i,j=1,..., N$. We assume that player $i$ wins the match against player $j$ with probability $p_i(a_i,a_j)$ and is rewarded $\Delta_j(a_j)$ by player $j$. Thus, with probability $p_i(a_i,a_j)$, the assets are updated as follows 
\begin{eqnarray}
a_i\rightarrow a_i+\Delta_j(a_j),\quad a_j\rightarrow a_j-\Delta_j(a_j).
\end{eqnarray} 
In turn, player $j$ wins the match with probability $p_j(a_i,a_j)=1-p_i(a_i,a_j)$ and then is rewarded $\Delta_i(a_i)$ by player $i$. Hence, with probability $p_j(a_i,a_j)$, the assets are updated as follows
\begin{eqnarray}
a_i\rightarrow a_i-\Delta_i(a_i),\quad a_j\rightarrow a_j+\Delta_i(a_i).
\end{eqnarray}
We require $\Delta_i(\cdot):\mathbb{R}_+\rightarrow\mathbb{R}_+$ be continuous functions, such that players never go bankrupt; i.e., players never reach zero assets paying for a match. Examples of $\Delta_i(\cdot)$ will be given subsequently. 

%Fig.~\ref{fig:1} shows a schematics of the configuration space of the game and the flow of probability toward the point $(a_1,a_2)$.

%\begin{figure}[h!]\begin{center}
%\includegraphics[width=0.5\textwidth]{FIG_K.pdf}
%\caption{A schematics for the configuration space of the game and the backward flow of probability.}\label{fig:1}\end{center}
%\end{figure}

We formally consider a time axis to order the gambling events consistently, at a rate of one event per unit time.  Hence, time is a counter for the total number of matches.  The symbol $f_t(\cdot)$ stands for the normalized distribution of the asset $a$ over the entire community, at time $t$. We write the backward Kolmogorov equation for the game, where $f_t(a_i)f_t(a_j)$ is the probability density that a player with asset $a_i$ meets a player with asset $a_j$ for a match at time $t$
\begin{eqnarray}
\label{eq:Kol}
\frac{\partial}{\partial t}[f_t(a_i)f_t(a_j)]=-f_t(a_i)f_t(a_j)+p_i(a'_i,a'_j)f_t(a'_i)f_t(a'_j)+p_j(a''_i,a''_j)f_t(a''_i)f_t(a''_j),
\end{eqnarray}
where
\begin{eqnarray}
\label{eq:a1a2a3a4}
a'_i-\Delta_i(a'_i)=a_i,\quad a'_j+\Delta_i(a'_i)=a_j, \quad a''_i+\Delta_j(a''_j)=a_i,\quad a''_j-\Delta_j(a''_j)=a_j,
\end{eqnarray}
such that $(a'_i, a'_j)$ are $(a''_i, a''_j)$ are the assets possible for the players $i$ and $j$ at previous time.
We are interested in equilibrium distributions of $a$, satisfying Eq.~\eqref{eq:Kol}. Particularly, we look for fixed-point solutions that are differentiable with respect to $a_i$. The fixed-point equation resulting from Eq.~\eqref{eq:Kol} is 
\begin{eqnarray}
\label{eq:fpKol}
f(a_i)f(a_j)=p_i(a_i+\Delta'_i,a_j-\Delta'_i) f(a_i+\Delta'_i)f(a_j-\Delta'_i)+p_j(a_i-\Delta''_j,a_j+\Delta''_j) f(a_i-\Delta''_j) f(a_j+\Delta''_j),
\end{eqnarray}
where $\Delta'_j\equiv\Delta_j(a'_j)$ and $\Delta''_i\equiv\Delta_i(a''_i)$.  

%In fact, by solving Eqs.~\eqref{eq:a1a2a3a4}, we may consider $\Delta'_j$ and $\Delta''_i$ to be implicit functions of $a_i$; i.e., $\Delta'_j(a_i,a_j)=\Delta_j(a'_i,a'_j)$ and $\Delta''_i(a_i,a_j)=\Delta_i(a''_i,a''_j)$. 

\section{Fixed-point analysis of an iterated zero-sum game}
We discuss a community consisting on $N$ gamblers that meet in random pairs to bet their assets on a coin toss. Hence, $p_{i,j}(a_i,a_j)$ are constant functions, independent of $a_{i,j}$, and $p_i+p_j=1$. We assume that, to avoid bankruptcy, each gambler bets half of his assets; i.e., $\Delta_i(a_i)=a_i/2$.  We are interested in the long-term distribution of assets in large communities where $N \rightarrow \infty$. 

To solve Eq.~\eqref{eq:fpKol}, we consider $a_{i,j}$, $\Delta'_i$ and $\Delta''_j$ in the vicinity of zero. We expand $f(\cdot)$ in series around $(a_i, a_j)$, and then formally set $a_i=a_j=a$. The zeroth and first order expanssion yield identies. The second order expanssion yields 
\begin{eqnarray}
%(\Delta_i'^2p_i+\Delta_j''^2p_j)\left[
f(a)\frac{\partial^2f}{\partial a^2}-\left(\frac{\partial f}{\partial a}\right)^2
%\right]
=0.
\end{eqnarray}
The only normalizable solution of the above equation is $f(a)=A^{-1}e^{-a/A}$, with $A>0$. Note that  $A=\langle a\rangle$ is time-independent since assets change through a zero-sum game. This exponential solution is independent on the initial asset configuration, representing a fixed point of the asset distribution. It is straightforward to check that the negative exponential is a solution of Eq.~\eqref{eq:fpKol}, provided that $p_{i,j}$ are constant functions. Remarkably, the fixed-point asset distribution is independent of the payment functions $\Delta_{i,j}(\cdot)$.

To investigate the stability of the fixed point solution, we perturb the solution of Eq.~\eqref{eq:Kol} as follows 
\begin{eqnarray}
f_t(a)=\langle a \rangle^{-1}e^{-a/\langle a \rangle}+\delta\! f(a,t),
\end{eqnarray}
where $\delta\! f(a,t)\ll f(a), \forall a$, at any moment of time $t$ that we consider. We obtain a differential equation which is first order in $\delta\! f(\cdot)$ and zeroth order in $\Delta'(\cdot,\cdot)\delta\! f(\cdot)$ and $\Delta''(\cdot,\cdot)\delta\! f(\cdot)$. We set $a_1=a_2=a$ and obtain
\begin{eqnarray}
\frac{\partial}{\partial t}\delta\! f(a,t)=-\delta\! f(a,t).
\end{eqnarray}
Hence, near the fixed point, the solution of the Eq.~\eqref{eq:Kol} has the form
\begin{eqnarray}
f_t(a)=\langle a \rangle^{-1}e^{-a/\langle a \rangle}+\delta\! f(a,0)e^{-t},
\end{eqnarray}
indicating that $f(\cdot)$ is a stable fixed-point solution. We recall that time was introduced formally, by the assumption that all players gamble at the same constant rate. 

\section{Numerical results}
The numerical algorithm is brute force integration of the zero-sum iterated game. We considered a vector of assets with $N=10^5$ entries. We randomly chose pairs of players, carried out a match for each pair, recorded the changes in the assets and repeated the process $\mathcal{O}(N^2)$ times. This ensured enough matches and good mixing between the players.

\begin{figure}[h!]\begin{center}
\includegraphics[width=.9\textwidth]{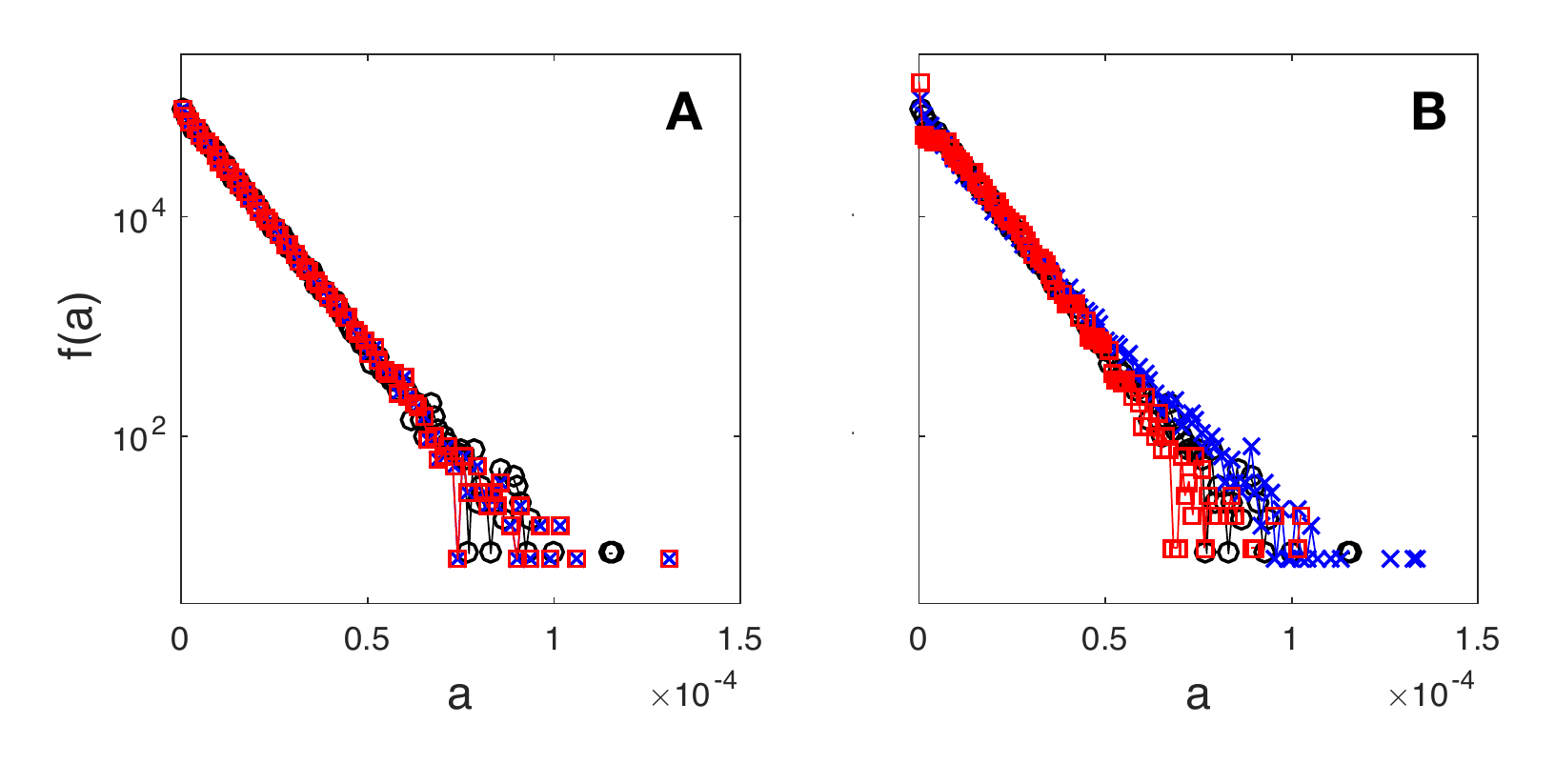}
\caption{Simulations of $f(a)$ versus $a$. The baseline game starts with $1/N$ assets for each player and uses $\Delta_i^{(1)}(a_i)=a_i/2$. We show results from variations of this game, to demonstrate properties of the fixed point solution $f(a)=\langle a \rangle^{-1}e^{-a/\langle a \rangle}$. Panel {\bf A} shows results emerging from different initial distributions $f_0(a)$, subsequently renormalized such that $\langle a\rangle=1/N$: all assets equal $1/N$ (black $\odot$), assets uniformly distributed in $[0,1]$ (blue $\times$) and assets normally distributed according to $\mathcal{N}(1/N,1/(5N))$ (red $\Box$). Panel {\bf B} shows results emerging from three different payment functions: $\Delta_i^{(1)}=a_i/2$ (black $\odot$), $\Delta_i^{(2)}=(0.25+\eta) a_i$, where $\eta$ is uniformly randomly distributed in $[0,1]$ (blue $\times$) and $\Delta_i^{(3)}=a_i\langle a\rangle/(a_i+\langle a\rangle)$ (red $\Box$).}\label{fig:2}\end{center}
\end{figure}

To check numerically that the asset distribution converges to a fixed point, we ran our game model for tree distinct initial condition distributions having the average $1/N$.  First, the initial distributions of assets was constant; i.e., assets were equal to $1/N$, for all players. Second, we chose the assets according to the uniform initial distribution $\mathcal{U}(0,1)$. Then, we renormalized the asset values such that the average of the initial condition distribution was $1/N$. Third, we chose the assets normally distributed according to $\mathcal{N}(1/N,1/(5N))$.  The simulation results are displayed in Fig.~\ref{fig:2}A. In particular, we note good collapse of the three curves originating from three different initial conditions, following the expected fixed-point exponential distribution.  

Next, we tested the dependence of the fixed point solution on the choice of $\Delta_i$. We started with all players having $a_i=1/N$ and consider three cases. First, we assumed that the loser pays half of his assets to the winner; i.e., $\Delta_i^{(1)}=a_i/2$. Second, we assumed that the loser pays a random fraction of his assets to the winner: $\Delta_i^{(2)}=(0.25+\eta) a_i$, where $\eta$ is chosen from $\mathcal{U}(0,1)$. Third, we assumed that the loser pays a fraction $\langle a\rangle/(a_i+\langle a\rangle)$ of his assets to the winner; i.e., $\Delta_i^{(3)}=a_i\langle a\rangle/(a_i+\langle a\rangle)$. Note that, in each of these three cases, the player paying $\Delta_i$ does not run out of assets because of payment. Numerical results show good collapse of the three simulations, all apparently converging to the fixed-point exponential distribution.

In conclusion, we investigated the dynamics of asset distribution in a large community where players meet one-on-one to play a zero-sum game. We referred to this zero sum game as a coin tossing game to stress out our assumption that the chance of winning does not depend on player's assets.  We found that, in the case where players can never go bankrupt, the asset distribution converges to the exponential distribution, irrespective of the betting procedure. This result was suggested by both analytical and numerical approaches.

%\bibliography{ComGame}

\begin{thebibliography}{10}

\bibitem{Mertens:1994ve}
J~F Mertens, S~Sorin, and S~Zamir.
\newblock {Repeated games}, 1994.

\bibitem{Hamilton:1981vw}
W~D Hamilton and R~Axelrod.
\newblock {The evolution of cooperation}.
\newblock {\em Science}, 211:1390--1396, 1981.

\bibitem{Kaznatcheev:2015}
Artem Kaznatcheev.
\newblock Short history of iterated prisoner's dilemma tournaments, March 2015.

\bibitem{Capraro:2013dz}
Valerio Capraro.
\newblock {A Model of Human Cooperation in Social Dilemmas}.
\newblock {\em PLoS One}, 8(8):e72427, August 2013.

\bibitem{Nowak:1993rm}
M~Nowak and K~Sigmund.
\newblock A strategy of win-stay, lose-shift that outperforms tit-for-tat in
  the prisoner's dilemma game.
\newblock {\em Nature}, 364(6432):56--8, Jul 1993.

\bibitem{Press:2012jh}
W~H Press and F~J Dyson.
\newblock {Iterated Prisoner's Dilemma contains strategies that dominate any
  evolutionary opponent}.
\newblock {\em P Natl Acad Sci USA}, 109(26):10409--10413, June 2012.

\bibitem{Hilbe:2013ip}
C~Hilbe, M~A Nowak, and K~Sigmund.
\newblock {Evolution of extortion in iterated prisoner's dilemma games}.
\newblock {\em P Natl Acad Sci USA}, 110(17):6913--8, 2013.

\bibitem{Wang:2016db}
Zhijian Wang, Yanran Zhou, Jaimie~W Lien, Jie Zheng, and Bin Xu.
\newblock Extortion can outperform generosity in the iterated prisoner's
  dilemma.
\newblock {\em Nat Commun}, 7:11125, 2016.

\bibitem{Rosenberg:2001ju}
Dinah Rosenberg and Sylvain Sorin.
\newblock {An operator approach to zero-sum repeated games}.
\newblock {\em Isr J Math}, 121(1):221--246, 2001.

\bibitem{Lehrer:2010iu}
Ehud Lehrer and Dinah Rosenberg.
\newblock {A note on the evaluation of information in zero-sum repeated games}.
\newblock {\em J Math Econ}, 46(4):393--399, July 2010.

\bibitem{Sorin:2011bj}
Sylvain Sorin.
\newblock {Zero-Sum Repeated Games: Recent Advances and~New~Links with
  Differential Games}.
\newblock {\em Dynamic Games and Applications}, 1(1):172--207, 2011.

\bibitem{Akian:2015}
M~Akian, S~Gaubert, and A~Hochart.
\newblock Ergodicity conditions for zero-sum games.
\newblock {\em Discrete Cont Dyn S}, 35(9):3901--3931, 2015.

\bibitem{ZiliottoB:2016}
B~Ziliotto.
\newblock Zero-sum repeated games: Counterexamples to the existence of the
  asymptotic value and the conjecture maxmin = lim v(n).
\newblock {\em Ann Probab}, 44(2):1107--1133, 2016.

\bibitem{Carmona:2016s}
G~Carmona and L~Carvalho.
\newblock Repeated two-person zero-sum games with unequal discounting and
  private monitoring.
\newblock {\em J Math Econ}, 63:131--138, 2016.

\end{thebibliography}

\end{document}